\documentstyle[12pt]{article}

\def\be{\begin{equation}}
\def\ee{\end{equation}}
\def\bdi{\begin{displaymath}}
\def\edi{\end{displaymath}}
\def\br{\begin{eqnarray}}
\def\er{\end{eqnarray}}

\def\u2{\mid u\mid^2}

\def\ra{\rightarrow}

\def\RR{{\rm I\kern-.1567em R}}                              
 \def\CC{{\rm C\kern-4.7pt                                    
 \vrule height 7.7pt width 0.4pt depth -0.5pt \phantom {.}}} 
 \def\ZZ{{\sf Z\kern-4.5pt Z}}                                

\begin{document}

\begin{titlepage}
\vspace*{-2 cm}
\noindent

\vskip 1cm
\begin{center}
{\Large\bf Knot soliton models, submodels, and their symmetries}

\vglue 1  true cm
C. Adam$^{1a}$ and  J. S\'anchez-Guill\'en$^{1b}$

\vspace{1 cm}

$^1${\footnotesize Departamento de F\'\i sica de Part\'\i culas,\\
Facultad de F\'\i sica,
Universidad de Santiago, \, \, and \\
Instituto Galego de Fisica de Altas Enerxias (IGFAE) \\
E-15782 Santiago de Compostela, Spain} \\ ${}$ \\

\vspace{1 cm}

\medskip
\end{center}

\normalsize
\vskip 0.2cm

\begin{abstract}
For some non-linear field theories which allow for soliton solutions,
submodels with infinitely many conservation laws can be defined. Here we 
investigate the symmetries of the submodels, where in some
cases we find
a symmetry enhancement for the submodels, whereas in others we do not.
\end{abstract}

\vfill

$^a${\footnotesize adam@fpaxp1.usc.es} 

$^b${\footnotesize joaquin@fpaxp1.usc.es}

\end{titlepage}

\section{Introduction}

Non-linear field theories with a two-dimensional target space 
and base space $\RR \times \RR^d$ ($d+1$ dimensional space-time)
can give rise to point like (vortex like) solitons for $d=2$,
or to line like (knot like) solitons for $d=3$, provided that the
fields are required to approach a fixed, constant value at spatial 
infinity (e.g., to render the energy finite), 
compactifying thereby the base space $\RR^d$.
Especially, for $d=3$ some  models with knot solitons have 
received considerable
attention recently and, further, such models have applications both in 
condensed matter \cite{BFN1,Bab1}
and elementary particle physics \cite{FN2,FNW1}.
Here we concretely consider models where the target space 
is the two-sphere $S^2$. Their solitons can  
be classified by the 
homotopy groups $\pi_2 (S^2)=\ZZ$ (winding number, for vortex type solitons) 
and $\pi_3 (S^2) =\ZZ$ (Hopf index, for knot type solitons),
respectively. 

The fields of the theories may be parametrized by a three-component unit 
vector ${\bf n}: \, \RR \times \RR^d \to S^2$, ${\bf n}^2 =1$, or
via the stereographic projection
\be
{\bf n} = \frac{1}{1+ u\bar u} \, ( u+\bar u , -i ( u-\bar u ) , 
u\bar u -1 ) \, ,\quad 
u  = \frac{n_1 + i n_2}{1 - n_3}
\label{stereo}
\ee
by a complex scalar field $u$. 

All models which we study can be constructed from the two Lagrangian densities
\be \label{cp1}
{\cal L}_2 = \frac{\partial_\mu u \, \partial^\mu \bar u}{(1+ u\bar u)^2}
\ee
and
\be
 {\cal L}_4 = \frac{(\partial^\mu u \, \partial_\mu \bar u)^2 - (\partial^\mu
u \, \partial_\mu u)(\partial^\nu \bar u \, 
\partial_\nu \bar u)}{(1+u\bar u)^4} .
\ee
In two space dimensions we consider the Baby Skyrme model ${\cal L}_{\rm BS}
={\cal L}_2$, whereas in three space dimensions we will consider the
Faddeev--Niemi model \cite{Fad,FN1}    with Lagrangian
 \be \label{FN-L}
{\cal L}_{\rm FN} = {\cal L}_2 - \lambda {\cal L}_4
\ee
(here $\lambda $ is a dimensionful coupling constant), the Nicole model
\be \label{Ni-La}
{\cal L}_{\rm Ni}= ({\cal L}_2)^\frac{3}{2}
\ee
(for which the one known soliton solution was found by Nicole, \cite{Ni}),
and the AFZ  model
\be
{\cal L}_{\rm AFZ} = -({\cal L}_4)^\frac{3}{4} ,
\ee
for which infinitely many soliton solutions have been found by
Aratyn, Ferreira and Zimerman (=AFZ) \cite{AFZ1,AFZ2}.
All four models circumvent Derrick's theorem and allow for static soliton
solutions, either by being spatially scale invariant (the Baby Skyrme,
the Nicole, and the AFZ model), 
or by involving two terms with opposite scaling behaviour
(the Faddeev--Niemi model).  

All four models (Baby Skyrme, Faddeev--Niemi, AFZ and Nicole) have the same
target space $S^2$, 
therefore they have some common properties. For instance,
all Lagrangians are invariant under modular transformations
\be
u \; \ra \; \frac{au + b}{-\bar b u +\bar a} \, , \quad a\bar a + b\bar b =1.
\ee
Furthermore, the same area-preserving diffeomorphisms on the target space $S^2$
can be defined for all models, but this does not imply
that they are symmetries for all four field theories. In fact, only the
AFZ model has the area-preserving diffeomorphisms as symmetries \cite{BF1,FR1}.
For the other three models the generators of the area-preserving
diffeomorphisms do not generate
symmetries and the corresponding Noether currents are not conserved. 
However, it was realized in the study of higher-dimensional integrability
within the generalization of the zero curvature
representation, \cite{AFSG}, that these Noether
currents {\em are} conserved for submodels of all three models
defined by the additional condition
\be \label{eik-eq}
\partial^\mu u\partial_\mu u =0 ,
\ee
i.e., the complex eikonal equation. Therefore, these submodels have
infinitely many conserved charges. On the other hand, their symmetries 
have to be determined independently, because the complex eikonal
equation is not of the Euler--Lagrange type, i.e., it does not follow
from an action, and the Noether theorem
does not apply to the submodels. This symmetry determination
is the main purpose of our talk. 

In Section 2 we give a very brief survey of the issue of integrability in
higher dimensions and of the resulting infinitely many conservation laws
(i.e., conserved currents). In Section 3 we introduce a general class
of Lagrangians (to which, of course, all models mentioned above belong) 
which provide a Lagrangian realization of the infinitely
many conserved currents of Section 2. Further, we explain the geometric
significance of these currents and their conservation.  
In Section 4 we briefly investigate the symmetries of the static equations
of motion (which are the relevant ones for solitons) for the three submodels
(of the Baby Skyrme, Faddeev--Niemi and Nicole models). Section 5 contains
our conclusions.

\section{Brief survey of integrability in higher dimensions}

In \cite{AFSG} a generalization of the zero curvature condition 
of Zakharov and Shabat in
1+1 dimensional integrable models was introduced in order to extend the 
concept of integrability to field theories in higher dimensions.
In its original formulation, this condition was a zero curvature in
a generalized loop space which leads to very non-local expressions when 
re-expressed in terms of fields over ordinary space-time. In the same
paper, however, a {\em local} condition realizing this generalized zero
curvature condition was given, which we want to describe briefly here. We
choose a non-semisimple Lie algebra $\tilde {\cal G}$ which is the direct sum
of a (possibly, but not necessarily semi-simple) Lie algebra ${\cal G}$ and
an abelian ideal ${\cal P}$, i.e.,
\be
\tilde {\cal G} = {\cal G} + {\cal P}
\ee
where ${\cal P}$ may, e.g., be a  (in general, reducible)
representation of ${\cal G}$,
in which case we have
\br
[T^a ,T^b ] = f^{abc} T^c & ,& [T^a, P^n ] = R^{mn}(T^a) P^m \nonumber \\  
{[} P^m , P^n {]} = 0 & , & T^a \in {\cal G} \, ,\quad P^m \in {\cal P}
\er
and $R^{mn}(T^a)$ are matrices in the representation ${\cal P}$.
Further, we choose a flat connection $A_\mu \in {\cal G}$, i.e.,
\be 
\partial_\mu A_\nu - \partial_\nu A_\mu + [ A_\mu , A_\nu ] =0
\quad \Rightarrow \quad A_\mu = g^{-1} \partial_\mu g
\ee
where $g \in {\rm \bf G}$ and ${\rm \bf G}$ is the Lie group of which
${\cal G}$ is the Lie algebra. Finally, we need a covariantly conserved,
vector-valued element of the abelian ideal, $B_\mu \in {\cal P}$, i.e., 
\be
\partial^\mu B_\mu + [A^\mu , B_\mu ] =0,
\ee
then there exist the conserved currents
\be
J_\mu \equiv g B_\mu g^{-1} \, ,  \qquad \partial^\mu J_\mu =0
\ee
as may be checked easily.  If this construction holds for
${\rm dim} ({\cal P}) =\infty$ then we have infinitly many conserved currents.

For our purposes we now specialize to  
${\rm\bf G}= SU(2)$ and choose as the group element $g$ a fixed, given
function of the field $u: \RR^d \times \RR \to \CC$ and its complex conjugate,
\be
g=g(u , \bar u) \in SU(2).
\ee
Essentially, $g$ takes values on the equatorial two-sphere contained within 
$SU(2)$ when $u$
takes values in $\CC$ (for an explicit expression see \cite{AFSG}).   
The representations $P^m$ are now
just the standard representations $P^{(l,m)}$ of
$SU(2)$ where $l$ and $m$ are the angular momentum and magnetic quantum
numbers, respectively. Further we restrict to $m=\pm 1$, i.e., 
\be
B_\mu = \sum_l c_l B_\mu^l ,
\ee
\be
B^l_\mu = K_\mu P^{(l,1)} + \bar K_\mu P^{(l,-1)}
\ee
where the $c_l$ are arbitrary real constants (making the abelian ideal 
infinite-dimensional), and $K_\mu (u ,\bar u ,u_\mu ,\bar u_\mu )$ is
a given function of the field variables $u,\bar u$ and its first derivatives
(in principle, also of higher derivatives, but we do not consider 
this possibility here). 
Here and below we use the notation $\partial_\mu u \equiv u_\mu$.
Different choices for $K_\mu$ correspond to different
field theories, as we shall see. Further, $K_\mu$ has to 
obey the reality condition
\be \label{re-cond}
\Im (\bar u_\mu K^\mu ) =0.
\ee
For the so chosen $B^l_\mu$, the corresponding currents 
$J_\mu^l = g B^l_\mu g^{-1}$ are equivalent to the currents
\be \label{J^G}
J_\mu^G = i(K_\mu G_u - \bar K_\mu G_{\bar u})
\ee
for an arbitrary {\em real} function $G(u , \bar u)$ ($G_u \equiv
\partial_u G$), see \cite{BF1}. 
If all $J_\mu^l$ are conserved, then $J_\mu^G$ is conserved for
arbitrary functions $G$. In the next Section, we shall present a Lagrangian
realization of these integrability concepts.  

\section{Lagrangian realization of conserved currents}

We introduce the class of Lagrangian densities
\be \label{g-lan}
{\cal L} (u ,\bar u ,u_\mu ,\bar u_\mu ) = F(a,b,c)   
\ee
where
\be
a=u\bar u \, ,\quad b=u_\mu \bar u^\mu \, ,\quad c= (u_\mu \bar u^\mu )^2
- u_\mu^2 \bar u_\nu^2 
\ee
and $F$ is at this moment an arbitrary real function of its arguments.
The phase
symmetry $u\ra e^{i\alpha} u$ for a constant $\alpha \in \RR$ holds by
construction. 
For the vector-valued function $K^\mu$ we choose
\be \label{k-mu}
K^\mu = f(a) \bar \Pi^\mu
\ee
where $f$ is a real function of its argument, and
$\Pi^\mu$ and $\bar \Pi^\mu$ are the conjugate four-momenta of
$u$ and $\bar u$, i.e. ($u_\mu \equiv \partial_\mu u$, $F_b \equiv
\partial_b F$, etc.),
\be
\Pi_\mu \equiv {\cal L}_{u^\mu} = \bar u^\mu F_b + 2 (u^\lambda \bar
u_\lambda \bar u_\mu - \bar u_\lambda^2 u_\mu )F_c.
\ee
$K^\mu$ in  Eq. (\ref{k-mu}) automatically obeys the reality condition
(\ref{re-cond}) for real Lagrangian densities. For the divergence
$\partial^\mu J^G_\mu$ of the current (\ref{J^G}) we find
\br \label{div-jg}
\partial^\mu J^G_\mu & = & if \left( [( M' \bar u G_u + G_{uu} ) u_\mu^2 
 - ( M' u G_{\bar u} + G_{\bar u\bar u}) \bar u_\mu^2 ] F_b  \right.
\nonumber \\
&& \left. + \, (uG_u - \bar u G_{\bar u}) [ M' (bF_b + 2 cF_c ) +F_a ] \right)
\er
where $M \equiv \ln f$,  
$M' \equiv \partial_a M$, and we used the equations of motion
\be \label{eom}
\partial^\mu \Pi_\mu ={\cal L}_u = \bar u F_a .
\ee
Now we want to study under which circumstances the divergence 
(\ref{div-jg}) vanishes (for a more detailed discussion we refer to 
Ref. \cite{ASG3}).
If no constraints are imposed neither on the Lagrangian 
nor on the allowed class of fields $u$, then we find the two equations for
$G$,
\be \label{geq1}
uG_u - \bar u G_{\bar u}=0,
\ee
and
\be \label{geq2}
M_a \bar u G_u + G_{uu} =0 \quad \Rightarrow \quad \partial_u [f(u\bar u)
G_u]=0 ,
\ee
with the solution 
\be
G_u = k\frac{\bar u}{f}
\ee
where $k$ is a real constant. The corresponding current $J^G_\mu$ is the
Noether current for the phase transformation
$u \ra e^{i \alpha}u $
which is a symmetry by construction.

Next we make the second term in (\ref{div-jg}) vanish by imposing on the
Lagrangian the condition
\be
M_a (bF_b + 2 cF_c ) + F_a =0
\ee
with the general solution
\be \label{sol-cha}
F(a,b,c)= \tilde F (\frac{b}{f},\frac{c}{f^2})
\ee
which has an interpretation in terms of the target space geometry.
In fact, introduce the two real target space coordinates $\xi^\alpha$ via
$u=\xi^1 +i\xi^2$ and the target space metric w.r.t. to the coordinates
$\xi^\alpha$ via
\br
g_{\alpha \beta } \equiv f^{-1} \delta_{\alpha \beta} 
& \Rightarrow &  {\rm det}(g_{\alpha \beta}) \equiv f^{-2} \\
\tilde \epsilon_{\alpha \beta}= f^{-1} \,
\epsilon_{\alpha \beta}  &&
\er
where $\epsilon_{\alpha \beta}$ is the usual antisymmetric symbol in two
dimensions.
Then the expressions on which
$\tilde F$ may depend can be written as
\br
\frac{b}{f}  &=& g_{\alpha \beta} (\xi)
\partial^\mu \xi^\alpha \partial_\mu \xi^\beta \\
\frac{c}{f^2}
&=& \tilde \epsilon_{\alpha \beta} \tilde \epsilon_{\gamma \delta}
\partial^\mu \xi^\alpha \partial_\mu \xi^\gamma \partial^\nu \xi^\beta
\partial_\nu \xi^\delta
\er
i.e., they depend on the target space metric and on the determinant of the 
target space metric, respectively. Let us point out here that all models of
Section 1 are of this type, i.e., ${\cal L}_2 = b/f$ and 
$ {\cal L}_4 = c/f^2$ for $ f=(1+a)^2$ (the target space metric of 
the two-sphere).

To make the first term in Eq. (\ref{div-jg}) vanish, as well, we may either
continue to impose Eq. (\ref{geq2}), which is solved by those $G$ which 
generate the target space isometries for the given target space metric
(i.e., the given function $f$). Or we may restrict the Lagrangian further by
imposing $F_b \equiv 0 \, \Rightarrow \, F=\tilde F(c/f^2)$. Then we have
no restriction on $G$ at all, and it follows that these unrestricted $G$ 
generate the area-preserving diffeomorphisms on target space. 
This is precisely the case
for the AFZ model.

Alternatively, we may make the first term in 
Eq. (\ref{div-jg}) vanish by imposing restrictions 
on the allowed field configurations $u$. In this case the currents $J^G_\mu$
are still the Noether currents of area-preserving diffeomorphisms, but these
transformations are no longer symmetry transformations of the pertinent
Lagrangians, in general. Concretely, we require that $u$ obeys the complex
eikonal equation (\ref{eik-eq}), which defines therefore submodels for all
models of the type (\ref{sol-cha}) with infinitely many conserved charges.

[Remark:  we might require, instead, that the field $u$ obeys the
(in general nonlinear) first order 
PDE which follows from the condition $F_b =0$ in cases
when this condition does not hold identically (i.e., for Lagrangians which do 
depend on the term $b=u^\mu \bar u_\mu$). This type of (``generalized'') 
integrability condition, which depends, however, on the chosen Lagrangian,
has been discussed in \cite{Wer3}, \cite{ASG3}.]

\section{Symmetries of the static equations}

Here we just want to present the results of the calculation of all geometric
symmetries (point symmetries) of the static equations, i.e., the static
equations of motion (e.o.m.) for the full models (Baby Skyrme, Nicole and
Faddeev--Niemi), and the static equations of motion plus the static
eikonal equation for the corresponding submodels (observe that the
static complex eikonal equation does have nontrivial solutions, in contrast to
its real counterpart, see, e.g., \cite{Ada1}). The method of prolongations
has been used for all symmetry calculations. Concretely, the symmetries of the
submodels are calculated by first calculating the on-shell symmetries
of the static eikonal equation. In a next step the on-shell symmetries of
the static second order equations are calculated, where the second order
equations consist of the equations of motion plus the prolongations of the
static eikonal equation (i.e., the second order equations that follow by 
acting with total derivatives on the complex eikonal equation). For the 
calculations we refer to \cite{ASG2}. Here we give a detailed discussion of
the results, which are displayed in Table 1.

\begin{table}

\begin{tabular}{||r|c|l|c||}
\hline
 & $\infty $ many & geometric &  solutions \\
 model & conserv. laws & symmetries &  known \\
\hline
Baby Skyrme & yes$^a$ & $C_2 \times SU(2) $  & yes \\
\hline
submodel & yes & $C_2 \times C_2 $ &  yes \\
\hline
Nicole  & no & $C_3 \times SU(2)$  & yes \\
\hline
submodel & yes & $C_3 \times SU(2)$ &  yes \\
\hline
Faddeev--Niemi & no & $E_3 \times SU(2)$  & yes$^b$ \\
\hline submodel & yes & $E_3 \times SU(2)$  & no \\
\hline 
\end{tabular} 

\caption{Some results for the three soliton models and
their submodels.  \newline
$C_d \; \ldots \; $ conformal group in $d$ dimensions. \newline
$E_d \; \ldots \; $ Euclidean group (translations and rotations) in $d$
dimensions. \newline
${}^a $due to the infinite-dimensional base space symmetries $C_2$. \newline
${}^b $known only numerically
}

\end{table}

For the Baby Skyrme model we find that the full static model has a
point symmetry group which is
a direct product of base space and target space
symmetries, where the group of base space symmetries is the conformal group
in two dimensions $C_2$, and the group of target space symmetries is the
group $SU(2)$. On the other hand, the submodel has the point symmetry group
$C_2 \times C_2$, i.e., the conformal group also in target space.
Therefore, the submodel has more symmetry in the case of the Baby Skyrme model,
although the additional symmetry is not related to the area-preserving 
diffeomorphisms. Further, there exist static solutions to the submodel. 
In fact, {\em all} soliton solutions of the Baby Skyrme model are, at the 
same time, also solutions of the submodel.

For the Nicole model we find that the group of point symmetries of the static 
e.o.m. is again a direct product of base space and target space symmetry
groups, where the base space symmetry group is $C_3$, the conformal group
in three dimensions, and the target space symmetry group is $SU(2)$.
Further, the static submodel has exactly the same symmetry group
$C_3 \times SU(2)$ as the full Nicole model. For the Nicole model only one
analytical soliton solution is known, but this solution solves the
static eikonal equation, as well, and is, therefore, also a solution of 
the submodel \cite{Ni,ASG2}.

For the Faddeev--Niemi model the situation is similar. Again, the 
static submodel has
exactly the same symmetries as the full static model, and the symmetry group is
a direct product of the Euclidean group in three dimensions $E_3$ in
base space (i.e., rotations and translations) and of the group $SU(2)$ in
target space. For the full Faddeev--Niemi model soliton solutions are
known only numerically \cite{FN2}, \cite{GH} - \cite{HiSa}. 
It is not known whether the submodel does or
does not have solutions.

\section{Conclusions}

In this talk we gave a brief survey of the generalized zero curvature
representation of \cite{AFSG},
which leads to a generalization of integrability to
higher-dimensional non-linear field theories. Then we introduced a
class of Lagrangian field theories parametrized by a complex field variable 
$u$, where this concept of integrability is realized by providing
infinitely many conserved currents either for the full theory or for 
the submodel defined by the eikonal equation $(\partial u)^2 =0$.
Further, these currents may be interpreted as Noether currents of
area-preserving diffeomorphisms on the target space where $u$ takes its 
values.
For some relevant models within this class of field theories,
which allow for soliton solutions, we then 
presented the results of a thorough analysis of the symmetries of 
their submodels. 

The general result for all cases is that the  area-preserving
diffeomorphisms are not symmetries of any eikonal submodel. Also, the 
three-dimensional  submodels of Faddeev--Niemi and Nicole have no additional
symmetries compared to the full theories.

The Baby Skyrme model is special, 
as the  restriction does have an intriguing
additional symmetry. This can be important as there is not much difference
of the solutions of the full model and the restriction, at least for the
static case. We remind that the method can be easily  extended to
include the time dependence.


\section*{Acknowledgments}

This research was partly supported by MCyT(Spain) and FEDER
(FPA2002-01161), Incentivos from Xunta de Galicia and the EC network
"EUCLID". Further, CA acknowledges support from the 
Austrian START award project FWF-Y-137-TEC  
and from the  FWF project P161 05 NO 5 of N.J. Mauser.


\end{document}